\documentclass[preprint2,trackchanges]{aastex701}

\usepackage{amsmath}

\newcommand\msun{$M_\odot$} 
\newcommand\logg{log~\textit{g}} 
\newcommand\cn{[C/N]} 
\newcommand\feh{[Fe/H]}

\newcommand\mgh{[Mg/H]}
\newcommand\mgfe{[Mg/Fe]}

\shorttitle{Checking It Twice}
\shortauthors{Roberts et. al.}

\begin{document}

\title{Checking It Twice: Using [C/N]-Masses and Asteroseismic Masses as a Diagnostic of Mass Loss and Transfer on the RGB}

\author[orcid=0000-0002-2854-5796, gname=John,sname=Roberts]{John D. Roberts}
\affiliation{Department of Astronomy, The Ohio State University \\
140 W 18th Ave \\
Columbus, OH 43210, USA}
\affiliation{Center for Cosmology and Astroparticle Physics, The Ohio State University\\
191 W Woodruff Ave \\
Columbus, OH, 43210, USA}
\email[show]{roberts.2158@osu.edu}  

\author[0000-0002-7549-7766]{Marc H. Pinsonneault}
\affiliation{Department of Astronomy, The Ohio State University \\
140 W 18th Ave \\
Columbus, OH 43210, USA}
\affiliation{Center for Cosmology and Astroparticle Physics, The Ohio State University\\
191 W Woodruff Ave \\
Columbus, OH, 43210, USA}
\email{pinsonneault.1@osu.edu}

\author[0000-0001-7258-1834]{Jennifer A. Johnson}
\affiliation{Department of Astronomy, The Ohio State University \\
140 W 18th Ave \\
Columbus, OH 43210, USA}
\affiliation{Center for Cosmology and Astroparticle Physics, The Ohio State University\\
191 W Woodruff Ave \\
Columbus, OH, 43210, USA}
\email{johnson.3064@osu.edu}

\author[0000-0003-0929-6541]{Madeline Howell}
\affiliation{Department of Astronomy, The Ohio State University \\
140 W 18th Ave \\
Columbus, OH 43210, USA}
\affiliation{Center for Cosmology and Astroparticle Physics, The Ohio State University\\
191 W Woodruff Ave \\
Columbus, OH, 43210, USA}
\email{howell.753@osu.edu}

\begin{abstract}

Red giants experience significant mass loss, but the mechanism is poorly understood. The surface \cn\ of red giants is correlated with birth mass, but not directly impacted by mass loss. Exploiting this, we compare asteroseismic masses of red giants with the same \cn\ and but different evolutionary states. We find bulk differences between stars at the beginning of the red giant branch and in the subsequent evolutionary phase, the red clump, providing a direct constraint on the strength of net RGB mass loss in field stars. We find that net mass loss decreases with metallicity and mass, matching recent studies for field giants, but contradicting expectations from the widely used Reimers' mass loss formula. We propose a mass- and metallicity-dependent Reimers' $\eta$ calibration that reproduces the empirical trends that we see. In addition, we identify 207 stars (3.33\% of our sample) that are clear outliers from their population in these birth mass bins, which we believe are likely candidates for mass transfer events. These stars do not show any obvious discrepancies in abundances or binary properties from their counterparts. This population should be accounted for in Galactic archaeological studies. Further follow-up is required to quantify their occurrence rate and origin.

\end{abstract}


\section{Introduction}\label{sec:intro}s

A star's mass is critical to determining its evolution and eventual fate. However, it has been known for decades that evolved red giants lose a large fraction of their mass \citep{IbenRood_1970}. This occurs in two distinct epochs: on the first-ascent red giant branch (RGB), before helium (He) ignition in the core, and on the asymptotic giant branch (AGB), after He core exhaustion. Our primary concern here is RGB mass loss. Red giant masses have taken on increased importance with the time domain revolution in astrophysics. Masses can be inferred for large numbers of red giants using asteroseismology \citep{ChaplinMiglio_2013}. Combined with composition,  now available from large spectroscopic surveys, mass can be used to infer ages, unlocking powerful diagnostics for Galactic archeaology \citep{Silva2018}. This reach can be greatly extended using asteroseismic data as a training set for larger data sets \citep{Martig_2016, Ness_2016}.  However, asteroseismology measures the \textit{current} mass, not the \textit{birth} mass, so an understanding of mass loss is crucial for reliable red giant ages. This can be further complicated for stars that experience interactions with companions. Stars with one or more companions can experience interactions leading to drastic changes in stellar mass, misleading inferred ages if this is not taken into account. Understanding the strength and likelihood of mass alteration, therefore, plays a large role in considering stellar populations and Galactic history.

RGB mass loss is well-known but, while plausibly linked to the low surface gravity of red giant stars, its mechanism remains poorly understood \citep{HolzerMacgregor_1985, Catelan_2009}. The commonly adopted ``Reimers' law'' assumes that a fraction of the star's luminosity goes into the gravitational potential energy of the lost mass \citep{Reimers_1975}. Other formulations are tied to physical mechanisms, such as turbulence, but are still dimensional scaling arguments \citep{SchroderCuntz_2005}. Mass loss prescriptions are typically calibrated in globular clusters \citep{Gratton_2010, Origlia_2014, Tailo_2020}. Given the old age of a globular cluster, the mass of stars evolved to the RGB is well predicted by theory. At their low metallicities and old ages, the color on the horizontal branch is strongly related to mass, allowing mass loss to be well-constrained in globular clusters. For younger or higher-metallicity samples, stars with a wide range of mass are found in the red clump (RC), making color alone insufficient. In this domain, however, asteroseismology can serve to constrain mass. Stellar seismic pulsations are related to the physical structure of the star \citep{Aerts_2010}. Through examining these pulsations, it is possible to determine the star's surface gravity and density \citep{Ulrich_1986, Brown_1991, KjeldsenBedding_1995}, which can then be used to determine mass. This method is not only precise, but is related to the current structure of the star, and thus the current mass. The use of asteroseismic masses in globular clusters has allowed mass loss measurements over a larger range of metallicities with improved precision \citep{Howell_2022, Howell_2024, Howell_2025}.

Extending this analysis outside of clusters proves difficult, however. To determine conclusively if a star's mass has changed, we require the ability to measure the mass at two different points in time. Asteroseismology allows measurements of current mass, but without clusters, constraining a star's birth mass is much more difficult. \citet{Li_2025} examined field stars filtering the sample using the age-velocity-dispersion relation and the RGB lower mass boundary. These methods restrict stars to be of similar age, and given the mass dependence of the main sequence, similar birth mass, effectively creating cluster-like samples of field stars. The results \citet{Li_2025} found in the field stars are in direct conflict with those of the globular clusters, showing that more examination of RGB mass loss in field stars is needed. We seek to provide measurements of RGB mass loss with greater resolution and at many different masses and metallicities using the \cn\ ratio. Stars undergo the First Dredge-Up (FDU), which mixes internal hydrogen-burning products into the convective envelope, and is proportional to mass \citep{Iben_1967}. CN-cycle processing leaves the stellar interior carbon-poor and nitrogen-rich, which means the surface \cn\ is correlated with the degree of mixing, and hence stellar mass. This method can be used to obtain mass estimates for stars up to 2-2.5 \msun~ (depending on the metallicity) \citep{Masseron_2015, Salaris_2015, Martig_2016, Roberts_2024}. The FDU occurs at a fixed point in a star's life: at the base of the RGB. This means that the \cn\ corresponds with a star's mass during FDU, effectively its birth mass. This relationship varies with \feh, due to differences in mixing efficiency and birth \cn, but post-FDU stars with the same \cn\ and \feh\ should all have had roughly the same mass. Barring additional effects such as extra-mixing in metal-poor stars \citep{Gratton_2000, Shetrone_2019}, \cn\ remains a window into a star's past mass. 

RGB mass loss is not the only effect that alters a star's mass; for stars with companions, more drastic changes can occur. Stars that are close enough can directly exchange mass through Roche-lobe overflows and other binary interactions, leading one star to lose mass to the other \citep{DeMarco_2017}. While these changes can be extreme, it is not always obvious that these stars have altered their mass unnaturally. Populations that have consistent masses, such as the Milky Way's high-alpha thick disk, make spotting these events simple. High-alpha stars are nearly uniformly old and low-mass \citep[e.g.]{Fuhrmann1998}. That makes the presence of young, high-mass stars in the population stick out quite clearly \citep[e.g.][]{Chiappini_2015, Lu_2025}, and they are often considered to be products of mass transfer events \citep[e.g.][]{Jofre_2016, Jofre_2023}. However, in the low-alpha thin disk, we expect stars to have a range of masses and ages at a given metallicity, so these mass-transfer products do not stick out the same way. These stars represent interesting subjects for understanding stellar interactions and also an abnormal population that must be considered appropriately in studies of stellar populations and Galactic archaeology. Thus, the ability to determine if stars have altered their mass strangely is of great value. The difference in ``measuring time'' between \cn\ and asteroseismology gives these two methods a powerful synergy. Using \cn, we have an estimate of the mass the star was born with. Using asteroseismology, we have a measurement of the current mass of the star. Comparing these two masses allows not only measurements of broad changes, but to locate stars that have drastically departed from the main population. We can check masses twice and determine which stars have been naughty and which have been nice.\footnote{Phrasing inspired by "Santa Claus is Comin' to Town", a song heard often during the writing of this paper.}

This idea has been used before to locate mass transfer candidates. \citet{Bufanda_2023} utilized APOKASC-2 \citep{apokasc2} to locate outliers in the \cn-mass trend and found that approximately 10\% of the stars did not follow the trend, with many of these having properties indicating binarity. \citet{Frazer_2025} compared age estimates from \mgfe-\feh\ abundances with asteroseismic ages in APOKASC-3 \citep{apokasc3}, a similar method but in a different domain. We seek to expand upon this precedent in two ways. First, we plan to locate mass-transfer candidates using a finer resolution of mass, taking advantage of the larger sample of stars in APOKASC-3. This will provide a clearer and more consistent picture of the abundance of these outliers in this domain. Second, we will explore a new avenue of using these different masses to measure RGB mass loss. Using \cn, we can build mono-mass samples of stars akin to clusters at any mass or metallicity, allowing us to explore mass loss in a sample of field stars with unprecedented resolution.

\section{Methods}\label{sec:data}

\subsection{Data Sources}
We use the data from the APOKASC-3 catalog \citep{apokasc3}, which combines spectral stellar parameters and abundances from APOGEE DR17 \citep{dr17} with asteroseismic data from the \textit{Kepler} mission \citep{Kepler}. The APOKASC-3 sample separates stars based on the number of independent asteroseismic parameter measurements. The gold sample, which contains 10,036 stars, has the most consistently measured stars and has the smallest mass uncertainties ($\sim4\%$). For this work, we utilize only these gold sample stars. Asteroseismic oscillation patterns also contain information about the core structure of a star, allowing for separation of first-ascent, shell-H burning RGB stars and core-He burning Red Clump (RC) stars \citep{Bedding2011}. We categorize the stars into two evolutionary categories: RGB stars, which have not experienced mass loss, and RC stars, which have. RGB stars are flagged as RGB in APOKASC-3 and have surface gravities between 2.5, to avoid deviations from asteroseismic relations at low gravities, and 3.3, to ensure FDU is complete. RC stars are flagged as RC in APOKASC-3 and have gravities between 2.2 and 3, to remove outliers. The \cn-mass relationship differs in high- and low-alpha stars due to differences in birth mixture \citep{Roberts_2024}. For this reason, we separate the low-alpha and high-alpha stars using the criteria from \citet{Weinberg_2017} updated to DR17 in \citet{Roberts_2024}. Finally, the ASPCAP errors in DR17 are underestimated \citep{apokasc3, Cao_2025}, so we adopt a minimum \feh\ error of 0.05 dex as done in \citet{apokasc3} and the recommended enhancement in [C/Fe] and [N/Fe] uncertainties from the appendix in \citet{Cao_2025}. With these conditions, our sample contains 6504 low-alpha stars (2539 RGB and 3975 RC) and 962 high-alpha stars (485 RGB and 477 RC).

\subsection{Analysis Methods}

In this paper, we use \cn\ and \feh\ to select RGB and RC stars of similar birth mass. We can then examine the differences between the current masses of the two populations as a signal of total RGB mass loss. For each birth mass bin, we find the median and median absolute deviation (MAD) of stellar mass for the RGB and RC separately. We use these values to find the difference in bulk populations, mass loss, and define outliers, most likely products of binary interactions.

For the low-alpha RGB and RC samples, we select stars within 0.05 dex of a given \cn\ and \feh. Within these windows, we calculate a preliminary median and MAD of the masses for the LRGB and RC individually. We mask 5$\sigma$ mass outliers using these values, then recalculate the median and MAD with the cleaned sample. If a window had fewer than 5 stars in either sample, we dropped it due to insufficient counts. For the high-alpha samples, the process is similar, but simpler. In the thick disk, there is a strong \feh-age correlation \citep{XiangRix_2022}, so at a given \feh\ there is a well-defined age and, correspondingly, \cn. Therefore, we select solely using \feh, and just mask \cn\ outliers within the window. \cn\ outliers were masked by fitting a line to the \cn-\feh\ distribution, determining the distribution of stars around this line, and removing stars that differ from this line by over 3 MAD. Metal-poor red clump stars are known to experience extra-mixing, lowering their \cn. For flagging stars as \cn\ outliers in the red clump, we first account for known depletion using the data from \citet{Shetrone_2019}, fitting their data points for \cn\ before extra-mixing and for the red clump, then constructing an interpolator for expected depletion at a given \feh. This brings the \cn\ values back to more closely matching the \cn\ values of the LRGB below \feh=-0.4 and allows us to use the same outlier criterion for the red clump.

\begin{figure}
    \centering
    \includegraphics[width=1\linewidth]{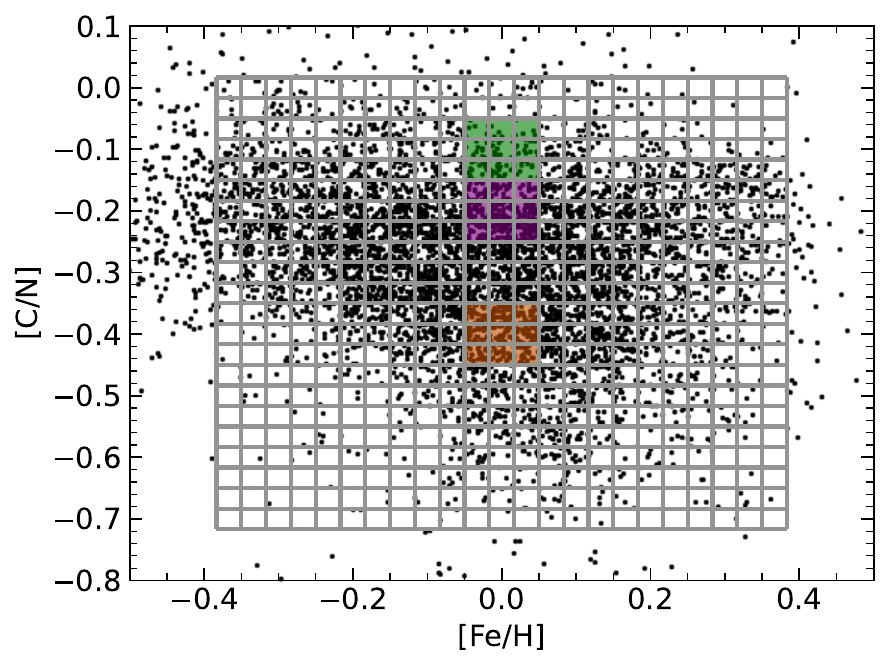}
    \caption{The RGB and RC stars in our sample, with lines showing the grid of \cn\ and \feh\ windows used in our analysis. The bins overlap, so a single window of 0.05 dex is a 3x3 grid of the shown squares. The green, purple, and orange regions show the windows used in the three panels of Figure \ref{fig:examplewindows}.}
    \label{fig:Windowbins}
\end{figure}

\begin{figure}
    \centering
    \includegraphics[width=\linewidth]{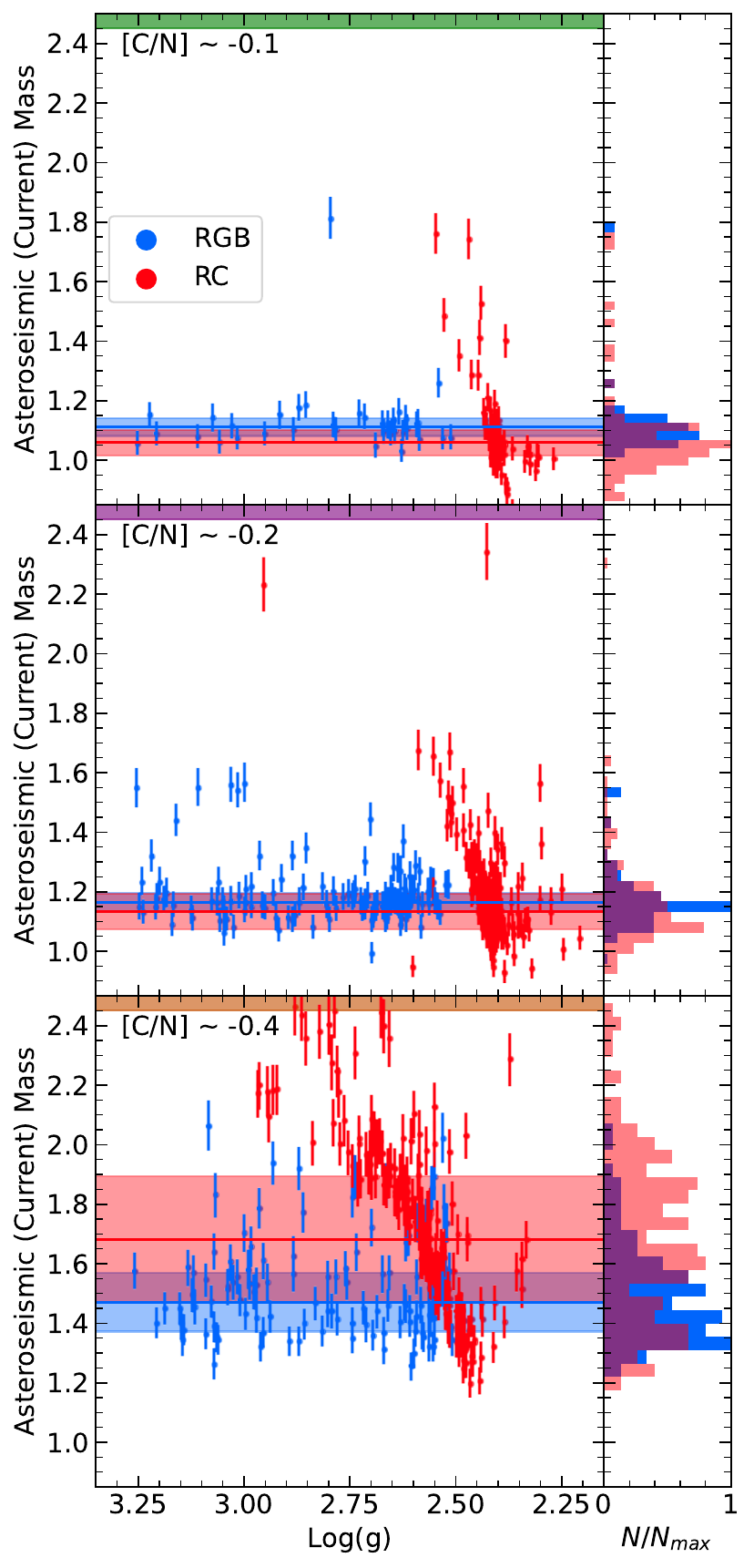}
    \caption{Mass versus \logg\ in \feh=0 windows at \cn$=-0.1,-0.2$ and $-0.4$. Blue points are LRGB, and red points are RC. The shaded bar at the top of each panel indicates the color of the shaded region in Figure \ref{fig:Windowbins} that is being plotted in that panel. The \logg\ dependence of mass on RC is expected because stars have similar radii on the red clump at different masses.}
    \label{fig:examplewindows}
\end{figure}

As a result of the mass dependence of RGB lifetime, our mass-metallicity domain is unevenly populated. In addition, as abundance data is uncertain, stars may be scattered into or out of windows erroneously. This poses a risk as abundance scatter can bias windows towards the most well-populated sample, obscuring the less populated domains. We measure the impact of these uncertainties by repeatedly shuffling the catalog abundances and repeating our analysis with the same asteroseismic masses. To shuffle abundances, we shifted the [C/N] and [Fe/H] of each star by a random value drawn from a normal distribution scaled to the uncertainty. We shuffled the abundances 99 times to generate 100 total median and median absolute deviation estimates for each window of \cn\ and \feh. For our final results, we take the median of these medians and MADs, which effectively weights the stars by their proximity to the bin. In combination with this, we allow our bins to overlap, which increases our resolution and smooths the trends we observe. For the low-alpha stars, our \cn\ bins extend from $-$0.7 to 0 (to cover the full range of well-populated \cn), and our \feh\ bins extend from $-0.3\overline{6}$ to $0.3\overline{6}$ (to cover the full range of well-populated \feh\ while avoiding extra-mixing). The individual windows are 0.1 dex wide, but we chose our step size to be one-third of this. This leaves us with 22 bins in \cn\ and 23 bins in \feh, creating a total of 506 windows. Figure \ref{fig:Windowbins} shows the \cn-\feh\ distribution of the low-alpha samples with our grid of windows overlaid, and the three windows used in Figure \ref{fig:examplewindows} shaded in different colors. Figure \ref{fig:examplewindows} shows the mass of the RGB and RC stars within these windows versus \logg. The grid for the high-alpha stars kept the same window and bin step sizes, but extended in \feh\ from -1 to 0.2, resulting in 37 total windows.

The shuffling process can demonstrate the impact of star-by-star uncertainties, but also amplifies biases resulting from the uneven shape of our distribution, as stars scatter into the window predominantly from one direction. To account for this effect, we forward modeled the effect by generating two synthetic samples. One was uniform in [C/N]-[Fe/H] space and extended 0.3 dex beyond the range of the windows in each dimension. The other was structured to mimic the [C/N]-[Fe/H] distribution of APOKASC-3. Approximate masses were assigned from the relationship from \citet{Roberts_2024}, adjusted in the high-mass domain to reproduce the overall [C/N]-mass distribution of APOKASC-3. After the masses were assigned, they were shuffled randomly, using mass errors of 4\%. These two samples were then subjected to the same analysis as above. The difference between the results is the measured impact of the shape of the distribution we are analyzing. These differences were fit to a second-order polynomial to smooth and used to correct the measured medians to more appropriate values. 

We note that the process of selecting similar mass stars using \cn\ loses potency in the high-mass, low-\cn\ domain due to two effects. First, at higher masses, the dependence of mixing on mass weakens \citep{Roberts_2024}. In fact, below a threshold \cn, the \cn-mass relationship becomes shallow enough that stars of a wide range of mass have the same \cn, and end up in the same window. This threshold varies with metallicity, but is close to \cn\ = $-0.38$ for solar metallicity stars. In addition, stellar lifetime in the RC depends more weakly on mass than lifetime on the RGB. For low-mass stars, the RGB and RC are both well-populated, as the stars evolve slowly enough to be readily observed. At the higher masses, though, RC stars become relatively more numerous as the RGB lifetime becomes short. The result is that in these higher-mass, lower [C/N] windows, the RC has significantly higher median mass than the RGB, as the corresponding high-mass RGB stars, which should also be in the window, are not observed. For both of these reasons, we limit our discussion to the [C/N]$<$-0.3 windows where these effects do not cause issues. 

\section{Results}\label{sec:results}

\subsection{Mass Loss on the RGB}

\begin{figure}
    \centering
    \includegraphics[width=1\linewidth]{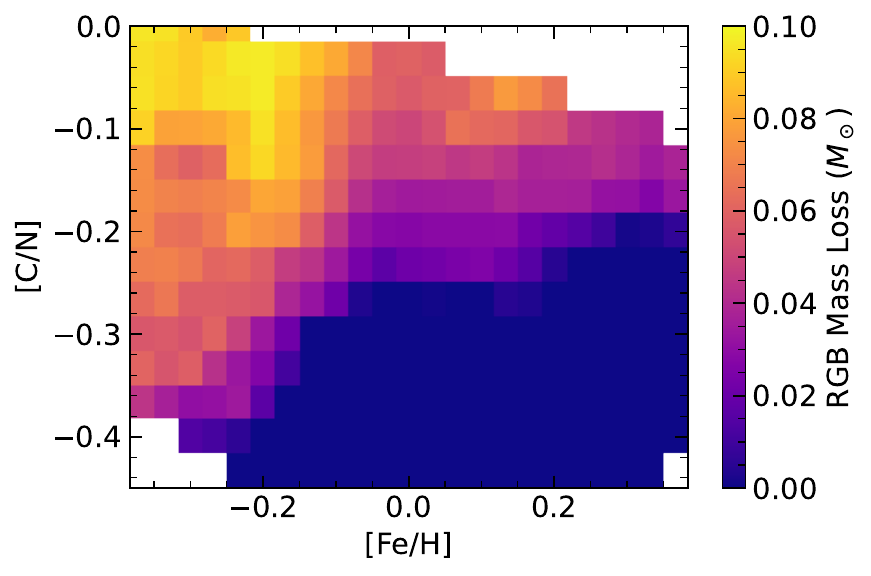}
    \caption{Net mass loss versus \cn\ and \feh. 
    In the low [C/N] - high [Fe/H] corner, the trend continues showing negative mass loss due to the earlier stated effects, so we adopt a minimum value of 0 for the color scale.}
    \label{fig:massloss}
\end{figure}

\begin{figure*}
    \centering
    \includegraphics[width=2\columnwidth]{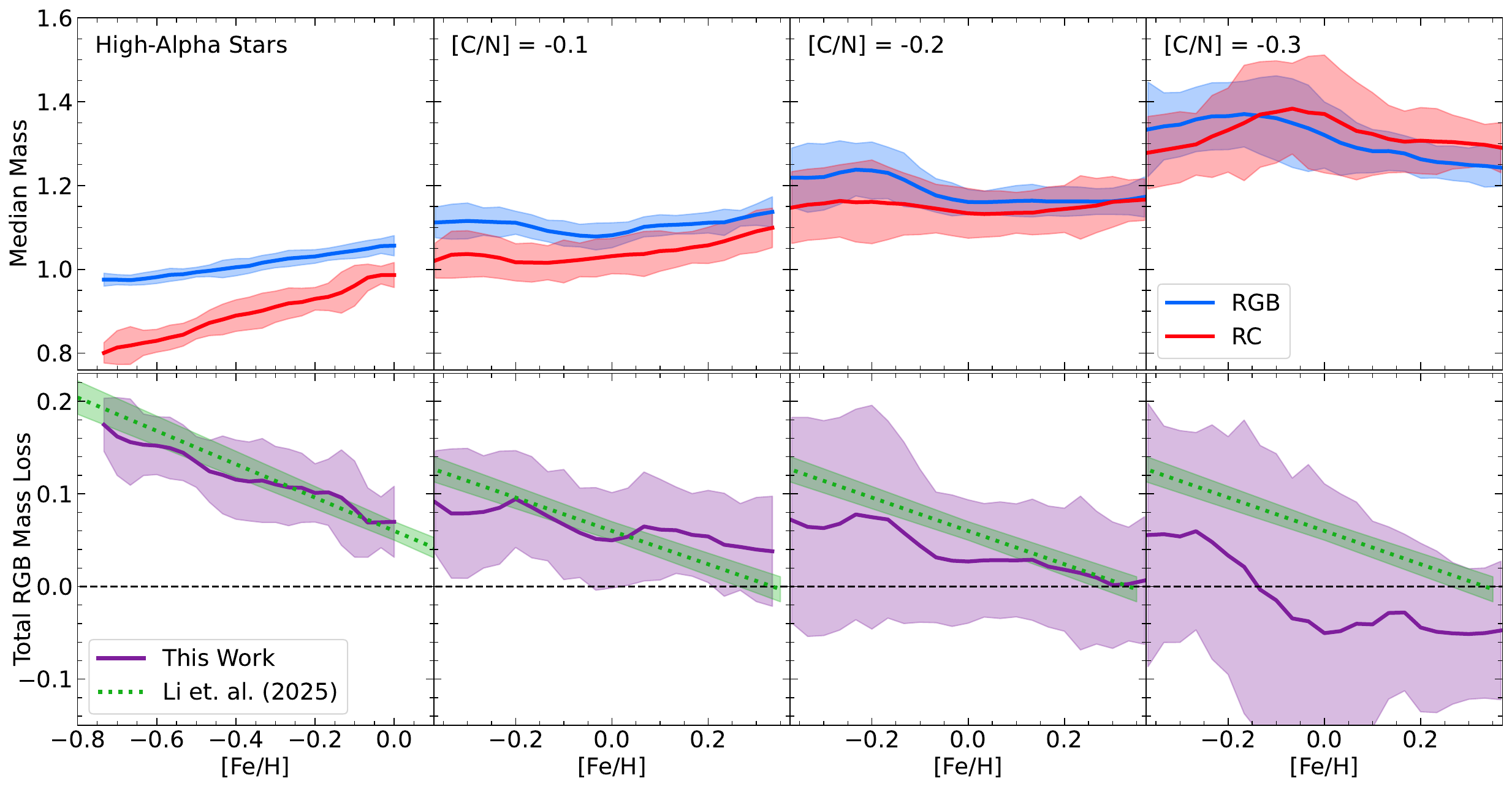}
    \caption{RGB (blue), RC (red) masses and mass differences (purple) versus \feh\ for the high alpha stars (left-most panels) and various \cn\ values. At lower \cn\ values, the uncertainty increases as the range in masses within the window grows. The trend from \citet{Li_2025} is shown in green.}
    \label{fig:panelcn}
\end{figure*}

By comparing the median mass of stars at the start and end of their RGB ascent, we can measure the amount of mass they lose during their RGB evolution. Figure \ref{fig:massloss} shows a map of the difference in mass\footnote{This mass difference is commonly referred to as integrated mass loss in the literature. Here we stick with the convention of calling it total RGB mass loss.} between the RGB and RC at fixed [Fe/H] and [C/N]. Figure \ref{fig:panelcn} shows the mass of the RGB, RC, and the mass loss trends at different \cn\ values. The bottom panels also show comparisons with \citet{Li_2025}. Qualitatively, these trends match our expectations regarding the mass dependence. Higher mass stars evolve faster, spend less time in their low-gravity, mass-shedding zone, and lose less mass. These trends also align with previous work, showing a decrease in mass loss with metallicity for field stars. This behavior opposes predictions from theoretical mass-loss prescriptions. Earlier studies of globular clusters show evidence of increased mass loss with higher metallicity, notably different than the trends seen in the field stars, despite using similar methods (See \citet{Li_2025}, Figure 7). Overall, our trends agree with those of \citet{Li_2025}, though showing a slightly shallower metallicity dependence. In the lower \cn\ bins, the trends are offset in magnitude; however, this is expected as their technique is valid for the field turnoff, while ours can be extended to younger ages and higher birth masses. 

\begin{figure*}
    \centering
    \includegraphics[width=2\columnwidth]{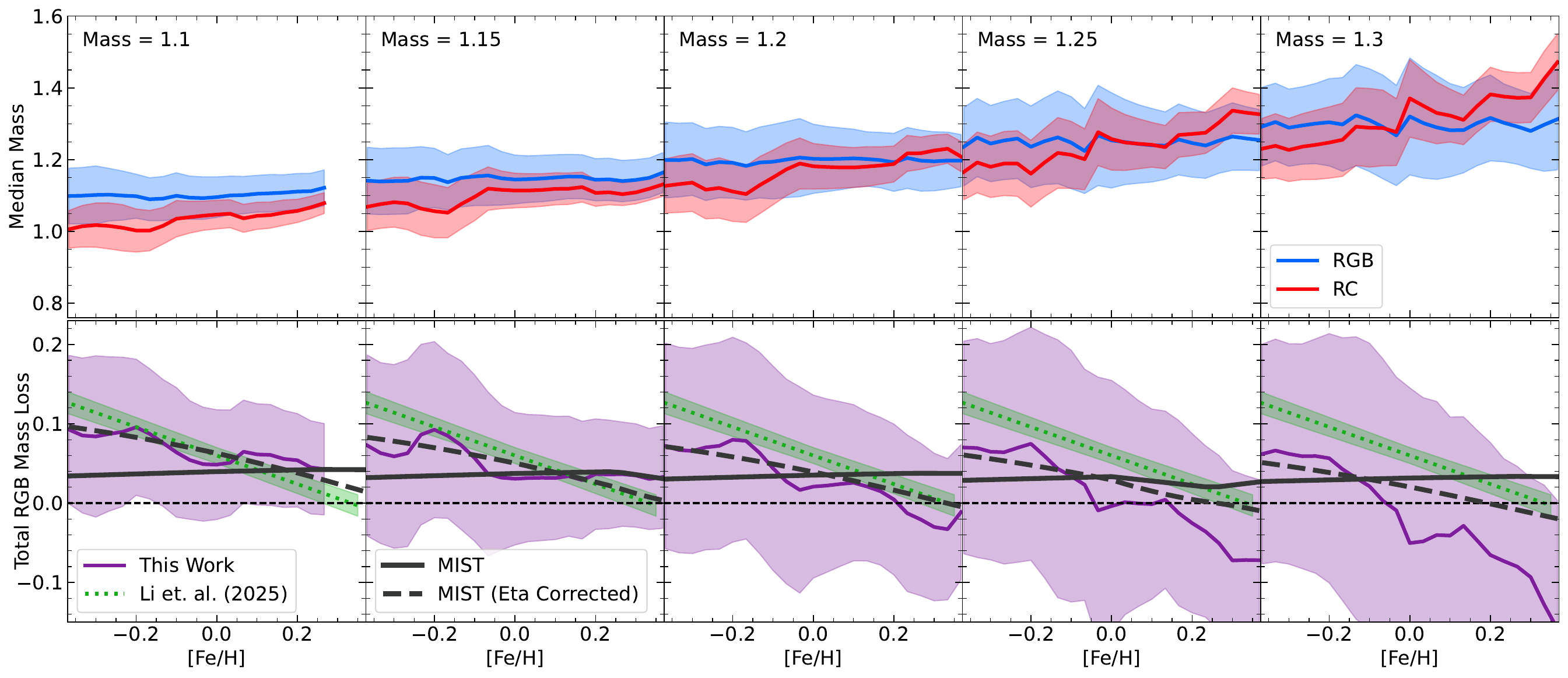}
    \caption{Same as Figure \ref{fig:panelcn}, but now using windows of fixed RGB mass. MIST mass loss values for the given mass are additionally plotted in the bottom panels with both the standard $\eta$ (solid line) and scaled by the variable $\eta$ in equation \ref{eq:eta} (dashed line). The primary change of the new $\eta$ is a metallicity-dependence as seen in the field stars.}
    \label{fig:panelmass}
\end{figure*}

Figure \ref{fig:panelmass} mimics Figure \ref{fig:panelcn}, but now each panel shows windows picked by their RGB mass rather than \cn, allowing for more direct comparisons with theoretical models. Each window with a mass within 2.5\% of a given mass value on the RGB was averaged together to create the lines shown. We have also now added the mass difference from theoretical expectations, taking from MIST evolutionary tracks \citep{MIST_0, MIST_1}, which uses a Reimers' prescription with $\eta=0.1$ for the RGB. The $\eta$ value is a free parameter used for calibrating the strength of mass loss (see equation \ref{eq:Reimer}). The MIST trends have a shallow, but positive correlation with metallicity, which disagrees with the results of both this work and \citet{Li_2025} and is only weakly dependent on mass. As can be seen, the Reimers' law with a constant $\eta$ does not reproduce the observed behavior of mass loss in the field giants. 

\begin{equation}\label{eq:Reimer}
    \dot{M}_\mathrm{Reimers} \propto \eta \frac{L_*R_*}{M_*}
\end{equation}

One possible solution to this disagreement would be to implement a variable Reimers' $\eta$ that brings existing theoretical trends into alignment. We use \feh\ and mass as dimensions, as those are two powerful factors of mass loss whose impact is not fully captured in the models, as shown in Figure \ref{fig:panelcn}. This variable $\eta$ does not provide a physical explanation of these disagreements, but can serve as a short-term fix to bring models into better agreement. We calculate the required $\eta$ through comparing the MIST tracks with the observed trends. Since feedback from the mass-loss would produce small perturbations in the luminosity, radius, and total mass of the star, a scaling of $\eta$ would, to first order, translate directly to a scaling of the final mass loss. Thus, we can calculate a ``corrected'' $\eta$ through taking the ratio of our observed mass loss and the MIST expected mass loss. Taking a minimum of 0 for the mass loss in the observations, we fit the ratios to find a new $\eta$ as a function of metallicity and mass. Our adjusted $\eta$ is provided in Equation \ref{eq:eta}. We note that this was calibrated in a range of $1.1<M_*/M_\odot<1.3$ and $-0.4<\mathrm{[Fe/H]}<0.4$, and should not be applied to stars far removed from this domain. This $\eta$ is intended to serve as a demonstration of the impact of metallicity and mass rather than a full replacement for $\eta$ in all stellar models. As $\eta$ is often calibrated for specific stellar populations, any implementation of this $\eta$ would, at minimum, need to reflect that it was calculated based on $\eta=0.1$ models, and should be scaled appropriately for populations that require stronger $\eta$ calibrations.

\begin{equation}\label{eq:eta}
\begin{aligned}
    \eta =  & ~~0.674 \\
    & -0.336 * \mathrm{[Fe/H]} \\
    & -0.469* \mathrm{Mass~(M_\odot)} 
\end{aligned}
\end{equation}

\subsection{Mass Transfer Events}

In addition to tracking the shift in mass of a population, we can also locate noticeable outliers in the mass-abundance domain. We interpolated the mass medians and MADs for the RGB and RC in \cn-\feh\ plane and assigned expected masses for each star based on their abundances. We used the \cn\ and \feh\ within their uncertainties that produced the mass closest to their asteroseismic mass, then measured the difference in terms of the MAD of the window. Using these differences, we separated the stars into two groups: Naughty (outliers differing by 3 $\sigma$ or more) and Nice (stars that are within 3 $\sigma$ of their expected mass). We remove from consideration stars whose \cn\ or \feh\ were far enough from the main distribution that the medians were not robustly defined, and are left with 207 stars on the Naughty list, 6199 stars on the Nice list, and 1060 stars that were outside the bounds of our interpolation. 

\begin{figure}
    \centering
    \includegraphics[width=1\linewidth]{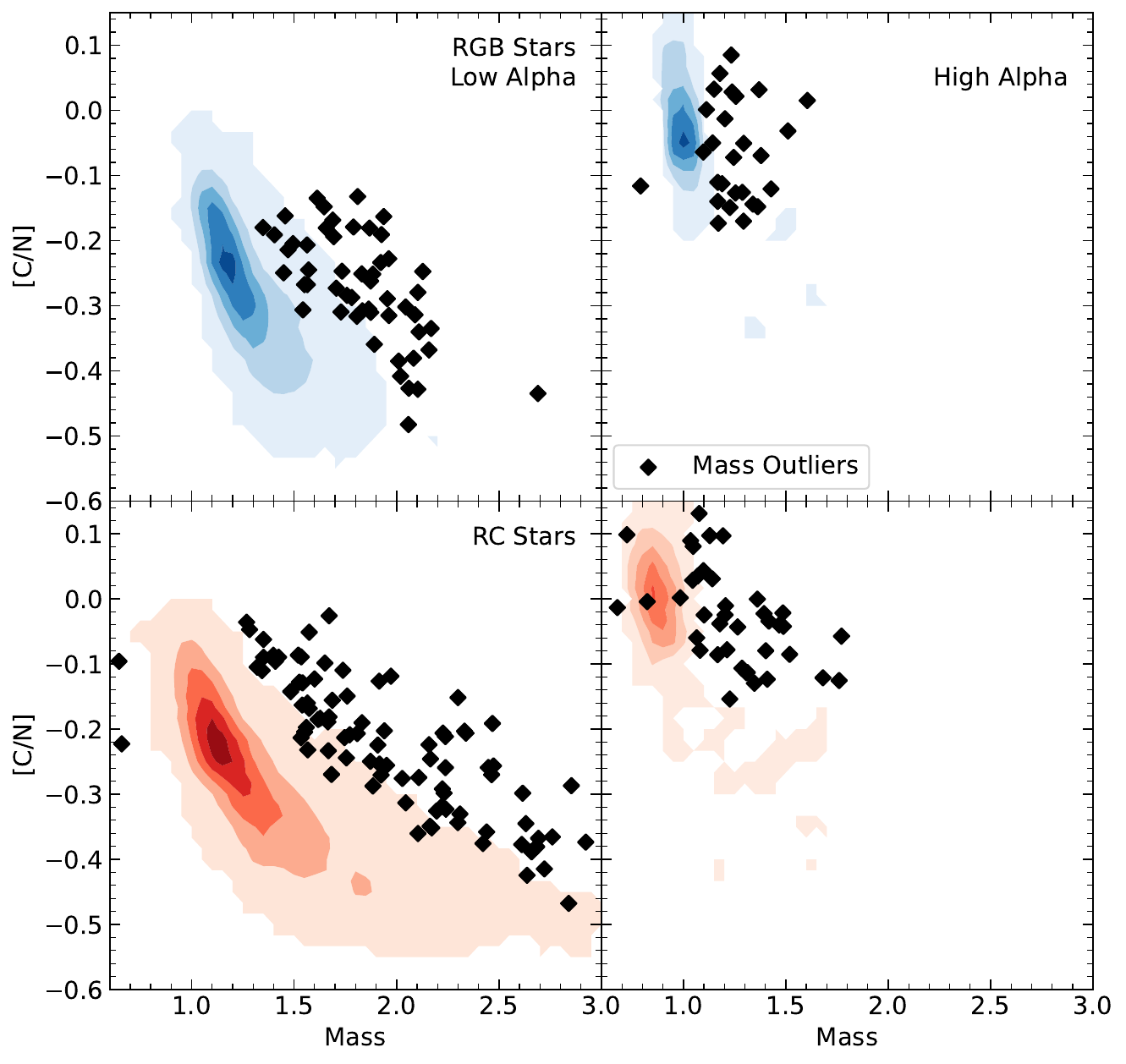}
    \caption{\cn\ versus mass for stars in our sample. Nice stars are plotted using smoothed contours, and Naughty stars are shown as black diamonds. The Naughty stars often fall outside the Nice distribution, but there are a few that have been flagged within the contours. These are stars that only appear as outliers when considering \feh\ as a dimension.}
    \label{fig:Naughty}
\end{figure}

Figure \ref{fig:Naughty} shows the Naughty and Nice samples in the \cn-mass domain. The fraction of stars flagged this way does not vary with metallicity in both the high- and low-alpha populations. 7.30\% of the high-alpha stars are flagged, though the low-alpha has only 2.56\%. This difference is not surprising as the distribution is much narrower in the high-alpha, leading to smaller MAD values. This difference should not be taken as evidence that outliers are more prevalent in the high-alpha domain, as it is they may just be easier to spot.

Previous works have used similar techniques to locate such outliers. \citet{Bufanda_2023} employed a similar technique using APOkASC-2, but grouped stars into 2 \feh\ bins, low and high. Compared with their method, we adopt much stricter $\sigma$ cuts yet still find stars close to the bulk distribution. We attribute this to the advantages of using a finer metallicity resolution to examine the stars. Overall, however, we grab fewer total stars, especially on the low mass side, where the errors are large compared to the discrepancies, due to our stricter requirements. 

\begin{figure}
    \centering
    \includegraphics[width=1\linewidth]{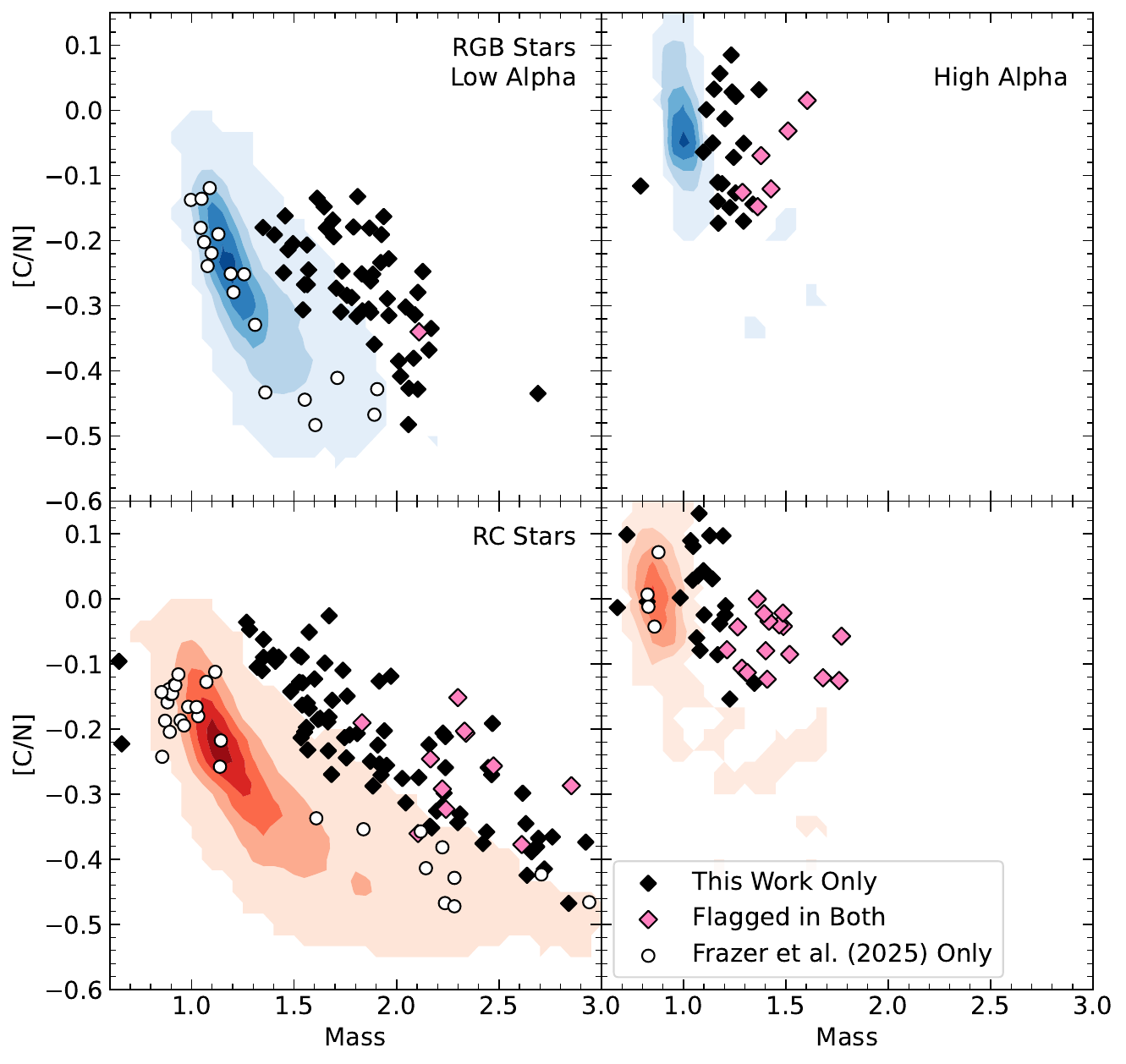}
    \caption{Figure \ref{fig:Naughty} but now showing the sample from \citet{Frazer_2025}. Diamonds have been flagged by our method, and non-black points have been flagged by theirs. We color the intersection of our samples differently for visual distinction. }
    \label{fig:Frazer}
\end{figure}

\citet{Frazer_2025} used APOKASC-3, but compared stars in the age domain, using \mgh and [Fe/Mg] assign ages. Since they used the same sample we did, we can directly compare stars flagged by the two different methods. Foremost, there is little overlap between our samples despite the conceptual similarity of our methods. Between the 213 stars we flag as anomalous and the 377 in \citet{Frazer_2025}, only 35 stars are flagged in both. Much of this difference comes from our boundary conditions. There are 290 stars they flagged that we do not because they either fall outside of our sample definitions for alpha-population or evolutionary state, or fall outside the \cn\ and \feh\ bounds where we can reliably determine a typical RGB and RC mass. This leaves 52 stars they identify as outliers that we do not, and 189 stars we flag that they do not (Figure \ref{fig:Frazer}). The differences between these two highlight the advantages of the different methods. The majority of stars only flagged by them are low-mass outliers, their mass donor candidates (MDCs). Meanwhile, we locate significantly more high mass outliers, their mass accretor candidates (MACs). This difference arises primarily from our choice of domain. Young high-mass stars can have large differences in mass, but relatively small age differences, and vice versa for old, low-mass stars. Thus, the age domain gives better separation on the lower mass side of the distribution, making it more effective for locating MDCs. Conversely, the mass domain is more effective at locating MACs. Given the complementary nature of our datasets and methods, we recommend using the union of our two sets for studies interested in exploring these outlier stars. Table \ref{tab:naughty} contains the IDs and parameters of our 207 flagged stars.

\begin{table*}[]
    \centering
        \begin{tabular}{c|c}
         Column Label & Contents \\
         \hline
         KIC, 2MASS\_ID, GAIAEDR3\_SOURCE\_ID & \textit{Kepler}, 2MASS and Gaia ID numbers \\
         Teff, Teff\_ERR & Effective Temperature and Error (K) \\ 
         Logg, Logg\_ERR & Surface Gravity and Error \\
         Mass, Mass\_ERR & Stellar Asteroseismic Mass and Error (\msun) \\
         EVstate & APOKASC3 Evolutionary State Flag \\
         FE\_H, FE\_H\_ERR & [Fe/H] and Error \\
         MG\_FE, MG\_FE\_ERR & [Mg/Fe] and Error \\
         C\_FE, C\_FE\_ERR & [C/Fe] and Error \\
         N\_FE, N\_FE\_ERR &  [N/Fe] and Error \\
         C\_N,  C\_N\_ERR & [C/N] and Error present in APOGEE DR17 \\
         C\_N\_Adjusted & [C/N] after correcting for extra mixing \\
         Mass\_window, MAD\_window & Expected Mass and Median Absolute Deviation (\msun) \\
         Mass\_window\_upper, Mass\_window\_lower & Upper and Lower possible expected mass (\msun) \\
         Difference\_Raw & Mass difference in factors of MAD using central values\\
         Difference & Difference calculated using closest points in error ranges \\
         \hline
        \end{tabular}
    \caption{Table of Outlier (Naughty) Stars. Empty cells in Mass window upper or Mass window lower indicate the star's errors brought it out of the interpolation range.}
    \label{tab:naughty}
\end{table*}

Following their example, we also explore whether or not these Naughty stars are peculiar in other ways beyond their asteroseismic mass. We examined the distributions of each elemental abundance given in APOGEE and found that the two distributions do not deviate in any axis. As binary interactions provide the simplest explanation of these outliers, we additionally examine typical diagnostics of binarity. APOGEE's VSCATTER measures the variation in radial velocity measurements between APOGEE observations, and GAIA's RUWE (Renormalized Unit Weight Error) measures deviations from standard astrometric fits that can be caused by binary companions \citep[e.g.][]{Berger_2020}. We do not find any evidence of these stars having a distinct VSCATTER distribution, and less than 10\% of the stars have a RUWE value above 1.4. This is similar to the \citet{Frazer_2025} sample, which showed no differences between mass transfer candidates and similar non-candidate stars in VSCATTER or RUWE. Some of these stars may be merger products, as \citet{Izzard_2018} found that high-alpha stars above 1.5 \msun are best explained as merger products. However, they also showed that for most models, 1.2-1.4 \msun stars are primarily still in observable binaries. Focusing on our high-alpha stars, most fall within this lower range, and so it is difficult to say these stars are primarily formed through accretion from a companion. Follow-up observations to determine if they are truly single, or examine core masses with asteroseismology to probe if they are the result of mergers, would be of great value in determining how these stars acquired their mass with their \cn.

\section{Summary and Conclusion}\label{sec:summ}

In this paper, we have used two different measures of mass, asteroseismology and \cn, to examine the change in mass stars experience on the RGB. Using APOKASC-3, we grouped stars into bins of similar birth mass using \cn\ and tracked changes in current mass with asteroseismology. With these measures, we have explored both mass loss and mass transfer events.

We provide empirical constraints on integrated mass loss at different masses and \feh\ for using field stars. We have provided additional evidence that mass loss does not follow a standard Reimers' convention in field stars and shows decreasing strength with metallicity, in conflict with globular clusters. To alleviate this, we calibrated a functional Reimers' $\eta$ that depends on metallicity and mass. This eta was based on the $\eta=0.1$ MIST isochrones and can be used to adjust mass loss rates in models to better match those of the field stars. We caution that this functional $\eta$ only serves to reduce the disparity in predictions, and does not carry an inherent physical meaning or interpretation. Further theoretical work on determining the nature of RGB mass loss is needed.

We also provide a sample of 207 stars that show disagreement between asteroseismic mass and typical \cn, which makes up 3.33\% of our total examinable sample. These stars do not show notable deviations from the other stars in their abundances or available diagnostics of binarity, which provides a lack of methods to detect them without asteroseismology. Similar stars are certainly present in large spectroscopic surveys, sitting undetected, and Galactic archaeological studies should take the presence of these stars into account. 

This method opens exciting future avenues. As TESS asteroseismic masses become available, the number of stars for which this analysis is possible will grow dramatically, though the increased uncertainties with respect to APOKASC may make confidently locating outlier stars more difficult. However, this could even be applied on a much larger scale without asteroseismology. In practice, the only necessary component for this is two independent mass estimates that have set ``encoding times.'' \cn\ will continue to be of high value in this respect as spectra are being acquired in large amounts for many stars, and it serves a valuable role as the ``look back'' metric. However, there are other ways to explore the current mass of a star. Spectroscopic gravities, for example, contain mass information and can also be acquired on a large scale. Though the precision may be lower than asteroseismology, the large increase in available stars could serve as a promising future direction of study. Relying on \cn, however, does place strict limits in the higher-mass regime. These limits, primarily arising from the weakening of \cn's mass dependence at higher masses, can be alleviated through higher precision spectroscopic abundances, though likely not eliminated. The lifetime constraints biasing the RGB to lower masses, too, could cause issues. This, however, could be corrected for through more nuanced characterization of the mass distributions, incorporating expected lifetime limits. 

\begin{acknowledgments}
We thank Polly Frazer for providing the list of their age outliers directly. We would like to thank the OSU Stars group for the fruitful discussions.

JDR, MHP, JJ, and MH were supported by NASA ADAP grant 80NSSC24K0637. JJ also acknowledges support from NSF grant AST-2307621.

Funding for the Sloan Digital Sky Survey IV has been provided by the Alfred P. Sloan Foundation, the U.S. Department of Energy Office of Science, and the Participating Institutions. SDSS acknowledges support and resources from the Center for High-Performance Computing at the University of Utah. The SDSS website is www.sdss4.org.

SDSS is managed by the Astrophysical Research Consortium for the Participating Institutions of the SDSS Collaboration including the Brazilian Participation Group, the Carnegie Institution for Science, Carnegie Mellon University, Center for Astrophysics | Harvard \& Smithsonian (CfA), the Chilean Participation Group, the French Participation Group, Instituto de Astrofísica de Canarias, The Johns Hopkins University, Kavli Institute for the Physics and Mathematics of the Universe (IPMU) / University of Tokyo, the Korean Participation Group, Lawrence Berkeley National Laboratory, Leibniz Institut für Astrophysik Potsdam (AIP), Max-Planck-Institut für Astronomie (MPIA Heidelberg), Max-Planck-Institut für Astrophysik (MPA Garching), Max-Planck-Institut für Extraterrestrische Physik (MPE), National Astronomical Observatories of China, New Mexico State University, New York University, University of Notre Dame, Observatório Nacional / MCTI, The Ohio State University, Pennsylvania State University, Shanghai Astronomical Observatory, United Kingdom Participation Group, Universidad Nacional Autónoma de México, University of Arizona, University of Colorado Boulder, University of Oxford, University of Portsmouth, University of Utah, University of Virginia, University of Washington, University of Wisconsin, Vanderbilt University, and Yale University.

\end{acknowledgments}

\begin{contribution}
J.D.R. was primarily responsible for the data analysis and writing of the manuscript. M.H.P. and J.A.J. collaborated closely on the analysis and interpretation, and M.H. provided expertise on the topic of mass loss and comments on the manuscript.
\end{contribution}

\facilities{APO, Kepler}

\software{astropy \citep{astropy2013,astropy2018,astropy2022},SciPy \citep{scipy}, NumPy \citep{numpy}, Matplotlib \citep{matplotlib}, pandas \citep{pandas}}

\bibliography{references}{}
\bibliographystyle{aasjournalv7}


\end{document}